# AUTHENTICATION SCHEMES USING POLYNOMIALS OVER NON-COMMUTATIVE RINGS

Maheswara Rao Valluri

School of Mathematical &Computing Sciences, College of Engineering, Science & Technology
Fiji National University, P.O.Box:7222, Derrick Campus, Suva, Fiji
`maheswara.valluri@fnu.ac.fj`

## ABSTRACT

*Authentication is a process by which an entity, which could be a person or intended computer, establishes its identity to another entity. In private and public computer networks including the Internet, authentication is commonly done through the use of logon passwords. Knowledge of the password is assumed to guarantee that the user is authentic. Internet business and many other transactions require a more stringent authentication process. The aim of this paper is to propose two authentication schemes based on general non-commutative rings. The key idea of the schemes is that for a given non-commutative ring; one can build polynomials on additive structure and takes them as underlying work structure. By doing so, one can implement authentication schemes, one of them being zero-knowledge interactive proofs of knowledge, on multiplicative structure of the ring. The security of the schemes is based on the intractability of the polynomial symmetrical decomposition problem over the given non-commutative ring.*

## KEYWORDS

*Authentication, Cryptography, Non-commutative rings, Polynomial rings, Protocols, &Security*

## 1. INTRODUCTION

The rapid world-wide development of electronic transaction simulates a strong demand for fast, secure and cheap public key cryptography (PKC). Public key cryptosystems are essential for electronic commerce or electronic banking transactions. They assure privacy of transaction as well as integrity of message for senders or receivers. The design of reliable public key cryptosystem presents a compendium challenging problems that have fascinated researchers in computer science, electrical engineering and mathematics alike for the past few decades and are certain to continue to do so.

Authentication is a process by which an entity, which could be a person or intended computer, establishes its identity to another entity. In private and public computer networks including the Internet, authentication is commonly done through the use of logon passwords. Knowledge of the password is assumed to guarantee that the user is authentic. Internet business and many other transactions require a more stringent authentication process.

Diffie and Hellman proposed in 1976 the public key cryptography [1]. They invented the concept of the encryption scheme and digital signature scheme based on public key. The trapdoor one-way functions play the key role in the idea of PKC. Today, most successful PKC schemes are based on the perceived difficulty of certain problems in particular large finite commutative rings. For example, see [1,3,4,6-8,10, 11, 13].

As addressed in [2], in order to enrich cryptography, there have been many attempts to develop alternative PKC based on different kinds of problems. Historically, some attempts were made for a cryptographic primitives construction using more complex algebraic systems instead of traditional finite cyclic groups or finite fields during the last decade. The originator in this trend was [7], where a proposition to use non-commutative groups and semigroups in session key agreement protocol was presented.

According to the author's knowledge, the first authentication scheme in an infinite non-commutative groups appeared in [14]. In [15], Cao et al. proposed a new Diffie-Hellman like key exchange protocol and ElGamallike cryptosystems using the polynomials over non-commutative rings. This enabled the authors of [15-17] to construct a public key cryptosystems and digital signatures. The Diffie-Hellman key agreement scheme also allows to construct two-pass challenge-response and iterated three-pass challenge-response authentication schemes.

In this paper, Diffie-Hellman and Fiat-Shamir like authentication schemes based on general non-commutative rings are proposed. The key idea of the schemes is that for given non-commutative ring, one can generate polynomials on additive structure and takes them as the underlying work structure. By doing so, one can implement authentication schemes over multiplicative structure of the rings. The security of the schemes is based on the intractability of the polynomial symmetrical decomposition problem over the given non-commutative ring.

The rest of the paper is organized as follows. In section 2, the well-known cryptographic assumptions over non-commutative groups are presented. In section 3,polynomials over an arbitrary non-commutative ring are defined and also necessary assumptions over non-commutative rings are presented. In section 4, Diffie-Hellman and Fiat-Shamir like authentication scheme based on underlying structure and assumptions are described, and also its security issues are discussed. Finally, concluding remarks are made in section 5.

## 2. CRYPTOGRAPHIC ASSUMPTIONS ON NON-COMMUTATIVE GROUPS

### 2.1 TWO WELL- KNOWN CRYPTOGRAPHIC ASSUMPTIONS

In a non-commutative group G, two elements x, y are conjugate, written x ~ y, if $y = z^{-1}xz$ for some z  G. Here z or $z^{-1}$ is called a conjugator. Over a non-commutative group G[12], one can define the following two cryptographic problems which are related to conjugacy.

-**Conjugator search problem (CSP):**Given (x, y)  G × G, find z  G such that $y = z^{-1}xz$ .

-**Decomposition problem (DP):** Given (x, y)  G × G, and S  G,find $z_1, z_2$  S such that $y = z_1xz_2.$

At present, it is believed that for general non-commutative group G, both of the above problems are difficult enough to be cryptographic assumptions.i.e., the CSP (DP, respectively) assumption says that CSP (DP, respectively) is intractable. More precisely, the CSP (DP, respectively) assumption states that there does not exist a probabilistic polynomial time algorithm which can solve CSP (DP, respectively) with non-negligible accuracy with respect to the problem scale, i.e., the number of input bits of the problem.

## 2.2 Symmetrical Decomposition and Computational Diffie-Hellman Assumptions over Non-Commutative Groups

Enlighted by the above problems, one can define the following cryptographic problems over a non-commutative group G.

**-Symmetrical Decomposition problem (SDP):** Given (x, y) ∈ G × G, and m, n ∈ Z, the set of integers, find z ∈ G such that y = $z^m$x $z^n$.

**-Generalized Symmetrical Decomposition Problem(GSDP):** Given (x, y) ∈ G × G, S ⊆ G and m, n ∈ Z, find z ∈ S such that y = $z^m$x $z^n$.

**-Computational Diffie-Hellman (CDH) Problem over Non-Commutative group G:** Compute $x^{z_1 z_2}$ (or) $x^{z_2 z_1}$ for given x, $x^{z_1}$ and $x^{z_2}$, where x ∈ G, and $z_1, z_2$ ∈ S, for S ⊆ G.

At present, there is no clue to solve this kind of CDH problem without extracting $z_1$ (or $z_2$) from x and $x^{z_1}$ (or $x^{z_2}$). Hence, the CDH assumption over G says that CDH problem over G is intractable. i.e., there is no probabilistic polynomial time algorithm which can solve CDH problem over G with non-negligible accuracy with respect to problem scale.

## 3. Building Blocks for Proposed Authentication Schemes

### 3.1 Integral co-efficient Ring Polynomials

Suppose that R is a ring with (R, +, 0) and (R, •, 1) as its additive abelian group and multiple non-abelian semigroup, respectively. Consider integral coefficient polynomials with ring assignment. At first, the notation of scale multiplication over R is already on hand. For $k \in Z_{>0}$ and r ∈ R,

$$(k)r \triangleq \underbrace{r + r + \cdots \ldots + r}_{k \text{ times}}.$$

For k =0, it is natural to define (k) r = 0.

**Remarks:** (1) (a)$r^m$.(b)$r^n$=(ab)$r^{m+n}$=(b)$r^n$.(a)$r^m$, ∀ a,b,m,n ∈ Z and ′r ∈ R.

(2) Note that in general,(a)r.(b)s ≠ (b)s.(a)r when r ≠ s, since the multiplication in R is non-commutative.

Now, suppose that $f(x) = a_0 + a_1 x + \ldots \ldots \ldots + a_n x^n \in Z_{>0}[x]$ is a given positive integral co-efficient polynomial. Then, one can assign this polynomial by using an element 'r' in R and eventually can obtain

$f(r) = \sum_{i=0}^{n}(a_i)r^i = a_0 + a_1 r + \ldots \ldots \ldots + a_n r^n$, which is an element in R.

Further, if r is regarded as a variable in R, then f(r) can be looked as polynomial about 'r'. The set of all this kind of polynomials, taking over all $f(x) \in Z_{>0}[x]$ can be looked the extension of $Z_{>0}$ with r, denoted by $Z_{>0}[r]$. It is called the set of 1-ary positive integral co-efficient R-polynomials.

Suppose that $f(r) = a_0 + a_1 r + \ldots \ldots \ldots + a_n r^n \in Z_{>0}[x]$ and $h(r) = b_0 + b_1 r + \ldots \ldots \ldots + b_m r^m \in Z_{>0}[x]$ for some n ≥ m. Then the following are held:

53

**Theorem1:** f(r).h(r) = h(r).f(r) ,   f(r), h(r) ∈ R

**Remark:** If r and s are two different variables in R, then f(r).h(s) ≠ h(s).f(r) in general.

### 3.2 FURTHER CRYPTOGRAPHIC ASSUMPTIONS ON NON-COMMUTATIVE RINGS

Let (R, +, •) be a non-commutative ring. For any a ∈ R, one can define a set $P_a$ ⊆ R by

$$P_a \triangleq \{f(a)/f(x) \in Z_{>0}[x]\}.$$

Then, consider the new versions of GSD and CDH problems over (R, •) with respect to its subset $P_a$, and name them as polynomial symmetrical decomposition (PSD) problem and polynomial Diffie-Hellman (PDH) problem-respectively.

**- Polynomial Symmetrical Decomposition (PSD) Problem over Non-Commutative Ring R:** Given(a, x, y) ∈ $R^3$ and m, n ∈ Z, find z ∈ $P_a$ such that y = $z^m x z^n$

**- Polynomial Diffie- Hellman (PDH) Problem over Non-Commutative Ring R:** Compute $x^{z_1 z_2}$ (or)$x^{z_2 z_1}$ for given x, $x^{z_1}$ and $x^{z_2}$, where a, x ∈ R, and $z_1, z_2$ ∈ $P_a$.

Accordingly, the PSD (PDH, respectively) cryptographic assumption says that PSD (PDH, respectively) problem over (R, •) is intractable. i.e., there does not exist probabilistic polynomial time algorithm which can solve PSD(PDH, respectively) problem over (R, •) with non-negligible accuracy with respect to problem scale.

## 4 .PROPOSED AUTHENTICATION SCHEMES

In this section, two public key authentication schemes (also called identification protocols) are described. The schemes meet necessary security compliance: they ensure the confidentiality of the private key, and the probability of finding another key that will behave like the private one is negligible. The security of the schemes relies on intractability of the polynomial symmetrical decomposition problem over the given non-commutative ring.

### 4.1 DIFFIE-HELLMAN LIKE AUTHENTICATION SCHEME FROM NON-COMMUTATIVE RINGS

The Diffie-Hellmanlike authentication scheme is a two-pass challenge-response scheme and is perfectly honest-verifier zero-knowledge.

This authentication scheme contains the following main steps: where Alice is the prover and Bob is the verifier.

**Initial Setup:** Suppose that the non-commutative ring (R,+, •) is the underlying work fundamental infrastructure and PSD problem is intractable on the non-commutative group (R, •). Choose two small integers m, n ∈ Z. Let H:R → $\mathcal{M}$ be a cryptographic hash function which maps from R to the message space $\mathcal{M}$. Then, the public parameters of the system would be the tuple<R, m, n, $\mathcal{M}$, H>.

**Key Generation:** First Alice selects two random elements p, q ∈ R and a random polynomial $f(x) \in Z_{>0}[x]$ such that $f(p)(\ne 0) \in$ R and then takes $f(p)$ as her private key, compute $y = f(p)^m q f(p)^n$ and publishes her public key $(p, q, y) \in R^3$.

**Authentication:** To begin authentication,

(a) Bob selects randomly another polynomial $h(x) \in Z_{>0}[x]$ such that $h(p) (\neq 0) \in R$ and computes the challenge $u = h(p)^m q h(p)^n$ then sends to Alice.
(b) Alice sends the response $w = H(f(p)^m u f(p)^n)$ to Bob, and Bob checks $w = H(h(p)^m y h(p)^n)$.

### 4.1.1 SECURITY ANALYSIS

In this subsection, the completeness, soundness and honest-verifier zero knowledge of the proposed scheme are examined.

**Completeness:** Assume that, at step (b), Alice sent $w'$. Then Bob accepts Alice's key if and only if $w' = H(h(p)^m y\, h(p)^n)$. The latter relation is equivalent to
$$w' = H(h(p)^m (f(p)^m q f(p)^n) h(p)^n) = H(f(p)^m (h(p)^m q h(p)^n) f(p)^n)$$
$$= H(f(p)^m\, u\, f(p)^n) = w. \text{ i.e. } w' = w.$$

**Soundness:** Assume a cheater $A'$ is accepted with non-negligible probability. This means that $A'$ can compute $H(h(p)^m y h(p)^n)$ with non-negligible probability. As H is supposed to be an ideal hash-function, $A'$ can compute an element $w \in R$ satisfying $H(w) = H(h(p)^m y h(p)^n)$ with non-negligible probability. Then, there are two possibilities: either $w = h(p)^m\, y h(p)^n$, which contradicts the hypothesis that the Diffie-Hellman like Polynomial Symmetric Decomposition Problem for q is hard [14], or $w \neq h(p)^m\, y\, h(p)^n$, which means that Cheater $A'$ and Bob are able to find a collision for H, contradicting the hypothesis that H is collision-free.

**Honest-Verifier Zero-knowledge:** Consider the probabilistic Turing machine defined as follows: It chooses random polynomial $h(p)$ using same drawing as the honest verifier, and outputs the instances $\bigl(h(p), H(h(p)^m\, yh(p)^n)\bigr)$. Then, the instances generated by this simulator follow the same probability distribution as the ones generated by the interactive pair (cheater $A'$, Bob).
For active attacks, the security is ensured by the Hash function H: if H is one-way, these attacks are ineffective.

### 4.2 FIAT-SHAMIR LIKE AUTHENTICATION SCHEME FROM NON-COMMUTATIVE RINGS

The Fiat-Shamir like authentication scheme [5] is a three-pass iterated challenge-response scheme. This scheme has to be repeated k times if one desires to reduce the probability of successful forgery to $\frac{1}{2^k}$.

This authentication scheme contains the following main steps: where Alice is the prover and Bob is the verifier.

**Initial Setup:** Suppose that the non-commutative ring (R,+, •) is the underlying work fundamental infrastructure and PSD problem is intractable on the non-commutative group (R, •). Choose two small integers m, n $\in$ Z. Let $\mathcal{M}$ be the message space. Then, the public parameters of the system would be the tuple <R, m, n, $\mathcal{M}$>.

**Key Generation:** First Alice selects two random elements p, q $\in$ R and a random polynomial $f(x) \in Z_{>0}[x]$ such that $f(p) (\neq 0) \in R$ and then takes $f(p)$ as her private key, computes $y = f(p)^m q f(p)^n$ and publishes public key $(p, q, y) \in R^3$.

**Authentication:** To begin authentication, Alice selects randomly another polynomial $h(x) \in Z_{>0}[x]$ such that $h(p)(\neq 0) \in R$ and then sends $u = h(p)^m y\, h(p)^n$, called the commitment, to Bob.

Bob chooses a random bit c and sends it as a challenge to Alice.

(a) If c = 0, then Alice sends $v(p) = h(p)$ to Bob. Bob aborts the authentication and rejects unless the equality $u = v(p)^m y v(p)^n$ is satisfied.

(b) If c = 1, then Alice sends $v(p) = f(p).h(p)$ to Bob. Bob aborts the authentication and rejects unless the equality $u = v(p)^m q v(p)^n$ is satisfied.

After k successful rounds, Bob accepts the authentication.

If Alice and Bob behave as specified, then the final check of each round leads
in step (a). to $v(p)^m\, y\, v(p)^n = h(p)^m\, y\, h(p)^n = u$, and in step (b). to $v(p)^m\, q\, v(p)^n = \big(f(p).h(p)\big)^m q \big(f(p).h(p)\big)^n = h(p)^m (f(p)^m\, q\, f(p)^n) h(p)^n = h(p)^m y\, h(p)^n$ = u.

### 4.2.1 SECURITY ANALYSIS

In this subsection, the completeness, soundness and zero knowledge of the proposed scheme are examined.

**Completeness:** In step (a), $u = h(p)^m y\, h(p)^n$. i.e. $u = v(p)^m y v(p)^n$.

In step (b), using $v(p) = f(p).h(p)$, one can find
$v(p)^m\, q\, v(p)^n = \big(f(p).h(p)\big)^m q \big(f(p).h(p)\big)^n = h(p)^m (f(p)^m\, q\, f(p)^n) h(p)^n = h(p)^m y\, h(p)^n$ = u.

Hence Bob accepts a correct answer at each repetition, so he accepts Alice proof of identity with probability 1.

**Soundness:** Suppose Alice does not possess the secret $f(p)$. Then, during any given round, she can provide only one of $v(p) = h(p)$ or $(p) = f(p).h(p)$. Therefore, an honest verifier, Bob will reject with probability in each round $\frac{1}{2}$ which implies an over all probability of $2^{-k}$ that a cheating prover will not be caught.

**Zero knowledge:** Consider the following probabilistic Turing machine $M'$:

-Step 1: $M'$ randomly selects a bit c and a polynomial $v(p)$.
-step2: For c= 0, $M'$ computes $u = v(p)^m y v(p)^n$; or c = 1, $M'$ computes $v(p)^m\, q\, v(p)^n = (f(p).h(p))^m q (f(p).h(p))^n = h(p)^m(f(p)^m\, q\, f(p)^n)h(p)^n = h(p)^m y\, h(p)^n = u$
-Step3: $M'$ initiates a protocol with Bob and sends u to Bob; Bob replies with the bit $c^1$ ;

-Step 4: For $c = c^1$, $M'$ outputs the triple $(u, c, v(p))$, otherwise it resets to step 1.

Since probability distribution of $h(p)$ in the authentication scheme is assumed to be right invariant, one can obtain the same probability distribution for the $v(p)'s$ generated by $M'$ as for Alice's ones. Moreover, since in case c =1 one can have $v(p)^m\, q\, v(p)^n = \big(f(p).h(p)\big)^m q \big(f(p).h(p)\big)^n$ $= h(p)^m (f(p)^m \, q \, f(p)^n) h(p)^n = h(p)^m y \, h(p)^n$ = u, using the same assumption, one can obtain the same probability distribution for the u's arising for c = 0 and those arising for c = 1. As a consequence, Bob cannot distinguish the two cases, and the probability to have $c = c^1$ is equal to ½. .

## 5. CONCLUSION

Recently, some promising authentication schemes have been proposed on non-commutative groups, such as braid groups. In this paper, a totally different methods have been used for describing Diffie-Hellman and Fiat-Shamir like authentication schemes based on general non-commutative ring. The key ideas behind schemes lies in the fact that one can take polynomials over the given non-commutative algebraic system as underlying work structure for constructing authentication schemes. The security of the proposed schemes is based on the intractability of polynomial symmetrical decomposition problem over the given non-commutative rings.

**Author:**

**Dr.MaheswaraRaoValluri**received the M.Sc., M.Phil., Ph.D. from Sri Krishnadevaraya University, Anantapur, A.P., India. He is currently working as an Assistant Professor, School of Mathematical & Computing Sciences, College of Engineering, Science, &Technology, Fiji National University, Derrick Campus, Suva, Fiji Island. His field of interest includes Cryptography, and Algebra.He is amember in International Association for Cryptologic Research and life member in Cryptology Research Society of India (CRSI), Andhra Pradesh Society of Mathematical Sciences (APSMS), and Ramanujan Mathematical Society (RMS), India.

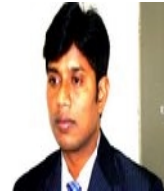